# Surface Equilibration Mechanism Controls the Stability of a Model Co-deposited Glass Mixture of Organic Semiconductors

*Shinian Cheng[1], Yejung Lee[1], Junguang Yu[2], Lian Yu[2], and M. D. Ediger*[*,1]*

[1]Department of Chemistry, University of Wisconsin-Madison, Madison, Wisconsin, 53706, USA

[2]School of Pharmacy, University of Wisconsin-Madison, Madison, Wisconsin, 53705, USA

**Corresponding Author**

*E-mail: ediger@chem.wisc.edu

## ABSTRACT

While previous work has identified the conditions for preparing ultrastable single-component organic glasses by physical vapor deposition (PVD), little is known about the stability of co-deposited mixtures. Here, we prepared binary PVD glasses of organic semiconductors, TPD (N,N'-Bis(3-methylphenyl)-N,N'-diphenylbenzidine) and m-MTDATA (4,4',4"-Tris[phenyl(m-tolyl)amino]triphenylamine), with 50:50 mass concentration over a wide range of substrate temperatures ($T_{sub}$). The enthalpy and kinetic stability are evaluated with differential scanning calorimetry and spectroscopic ellipsometry. Binary organic semiconductor glasses with exceptional thermodynamic and kinetic stability comparable to the most stable single-component organic glasses are obtained when deposited at $T_{sub}$=0.78-0.90$T_g$ (where $T_g$ is the conventional glass transition temperature). When deposited at 0.94$T_g$, the enthalpy of m-MTDATA/TPD glass equals that expected for the equilibrium liquid at that temperature. Thus, the surface equilibration mechanism previously advanced for single-component PVD glasses is also applicable for these co-deposited glasses. These results provide an avenue for designing high-performance organic electronic devices.

## TOC GRAPHICS

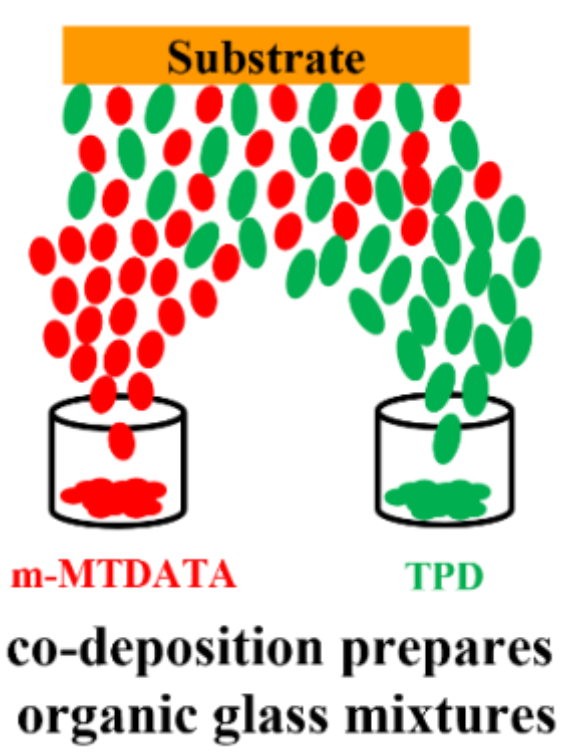

Glasses play a central role in modern technologies, including communications[1], pharmaceuticals[2], and organic electronics[3]. They are amorphous solids with macroscopic homogeneity and nearly unlimited compositional flexibility. These features make glasses preferable to crystalline materials in some applications, such as organic light emitting diode (OLED) displays. The active layers in OLEDs are organic semiconductor glasses. The macroscopic homogeneity of glasses ensures smooth surfaces and uniform performance in all pixels, while the compositional flexibility of the glassy matrix facilitates the preparation of well-mixed emissive layers with tunable composition. A fundamental challenge for glass materials is their long-time stability[4,5]. Due to their non-equilibrium nature, glasses can either physically age[6] over time or undergo crystallization if chemical degradation is prevented[7]. Both may lead to degradation of OLED device performance and reduce lifetime[8]. Therefore, it is practically important to produce glass materials with highly enhanced stability.

Recent studies have demonstrated that physical vapor deposition is an excellent technique to prepare glasses with exceptional thermodynamic and kinetic stability[9,10,11,12]. In addition, such ultrastable PVD glasses also exhibit high density[13], enhanced photostability[14,15], high resistance to crystallization[16], and high mechanical moduli[17,18]. These desirable properties cannot be obtained using other preparation techniques. It is hypothesized that surface mobility is responsible for the formation of these ultrastable glasses prepared from PVD[11]. The strongly enhanced mobility at the glass surface allows molecules to find low energy configurations before being buried by further deposition[11]. This surface equilibrium mechanism has been supported by theoretical work[19,20], computer simulations[21,22,23], and direct surface mobility measurements[24,25,26,27]. With a few exceptions[28,29,30], however, the study of the stability of PVD organic glasses has been limited to single component systems.

Some fundamental issues remain to be addressed regarding the stability of PVD glass mixtures. It is not clear whether a mixture should form an ultrastable glass, even when the two components individually form ultrastable glasses. Based on the knowledge from single-component PVD glasses, high mobility at the surface is the key for molecules to find low energy configurations and form ultrastable glasses. However, it may be impossible to find a proper deposition temperature, at which both components for co-deposition have high surface mobility simultaneously, especially when they have a large difference in $T_g$ values. In addition, even if the two pure components can have high enough surface mobility at a given temperature, immiscibility or the capability to form hydrogen bonds between components (which would lower surface mobility) may block formation of ultrastable mixtures. Very recent work reported that co-deposited organic semiconductor glasses of 8-hydroxyquinolinolato-lithium (Liq) and 4,7-Diphenyl-1,10-phenanthroline (BPhen) do not show ultrastable properties (*e.g.* higher density)[30].

Understanding the properties of PVD glass mixtures is important for technology. PVD is the standard route to prepare glassy layers of organic semiconductors in OLEDs and these layers are often mixtures. For example, the light-emitter layer is generally a glassy mixture of light emitting molecules dispersed in a host[31,32,33]. Recent studies have indicated that OLEDs prepared with ultrastable vapor-deposited glass layers show extended device lifetime[34,35]. Thus, it is an important goal to understand the physical mechanisms controlling the stability of vapor-deposited glass mixtures and to identify the deposition conditions producing highly stable multicomponent glasses.

To enrich our understanding of multicomponent PVD glasses, we co-deposited binary glasses of organic semiconductors: m-MTDATA and TPD, in a wide substrate temperature ($T_{sub}$) range. The two components selected can form ultrastable glasses as neat materials[36]. Differential scanning calorimetry and spectroscopic ellipsometry were applied to evaluate the enthalpies and kinetic stabilities of the PVD mixtures. In this work, we focused our attention on non-dilute mixtures with mass ratio near 50:50, as we anticipate that this is the regime in which ultrastable glass formation is most challenging. We found that the stability of co-deposited m-MTDATA/TPD glasses is controlled by $T_{sub}/T_g$ (where $T_g$ is the conventional glass transition temperature) in the same manner as single-component PVD organic glasses. When co-deposited at $T_{sub}$=0.78-0.90$T_g$, the most stable binary glasses are formed with an onset temperature being 5% higher than the conventional glass transition temperature, which is comparable to the most stable single-component organic glasses. Interestingly, the enthalpy of m-MTDATA/TPD glass deposited at 0.94$T_g$ equals that expected for the equilibrium liquid at that temperature. All these results are consistent with the surface equilibration mechanism previously advanced to understand single-component PVD glasses.

The DSC results in Figure 1A demonstrate that the co-deposited m-MTDATA and TPD (chemical structures shown in Figure 1) glass mixture with 50:50 mass ratio prepared at $T_{sub}$=300K is kinetically much more stable than the corresponding liquid-cooled glass. The results for the as-deposited sample obtained in the initial heating process are presented in pink. After the as-deposited glass is completely transformed into the liquid state, the same sample is cooled by 10K/min to form the corresponding liquid-cooled glass and then heated again yielding the gray data. As shown in Figure 1A, the devitrification process for both deposited and liquid-cooled glass is accompanied by a significant increase in heat capacity. The onset temperature ($T_{onset}$) where the as-deposited glass starts to transform is 361.3K, while the glass transition temperature ($T_g$) characterized using the mid-point convention for the corresponding liquid-cooled glass is 341.1K. This 20.2 K difference suggests that a much higher temperature is required for the vapor-deposited sample to disrupt its glassy molecular packing, a straightforward indication of higher kinetic stability for the co-deposited m-MTDATA/TPD glass. It should be emphasized that the as-deposited films are fully amorphous. The absence of crystalline material is confirmed by grazing-incidence wide-angle x-ray scattering (GIWAXS) and DSC measurements (see Figure S1 and S2).

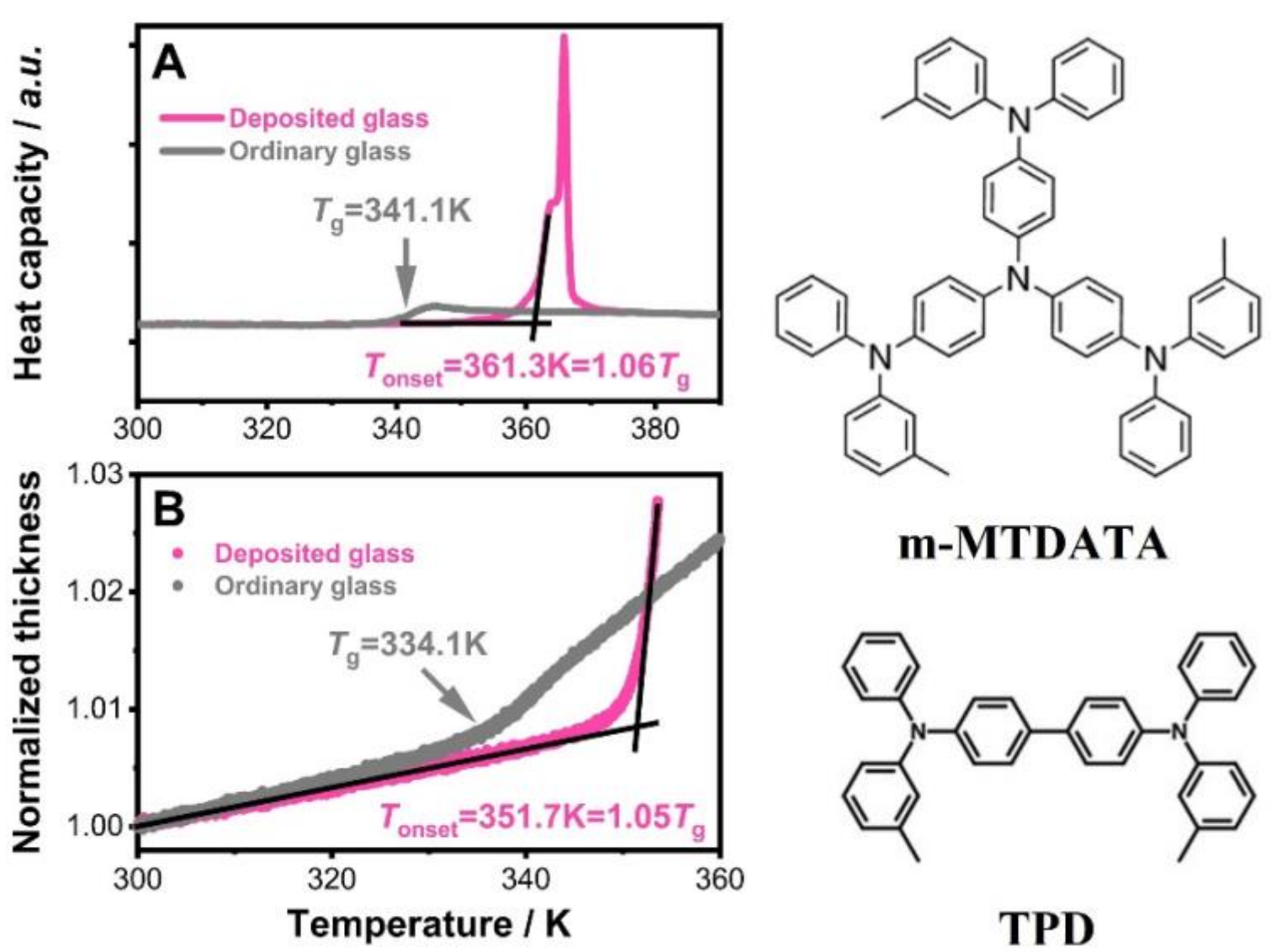

**Figure 1.** Temperature scanning experiments to determine the kinetic stability of co-deposited m-MTDATA/TPD glasses with mass ratio 50:50. The binary glasses were prepared at $T_{sub}$ = 300K with deposition rate 0.42±0.03nm/s. A) Heat capacity as a function of temperature determined from DSC measurements in the heating process with 10K/min.; B) Normalized film thickness as a function of temperature determined from ellipsometry ramping measurements at a heating rate of 1K/min. The film thickness is normalized at 300K.

One challenge in performing calorimetric measurements on thin vapor-deposited glasses is to introduce sufficient sample mass for good thermal signals. C. Rodríguez-Tinoco *et al.*[37] addressed this issue by using aluminum foil as the substrate and then folding the foil with deposited glass into a DSC pan. In this work, we used gold foil rather than aluminum because of its better thermal conductance. We co-deposited a 1200nm organic film onto a $1.8\times1.8cm^2$ gold foil; the foil mass was about 0.4mg while the sample mass was about 0.5mg. As shown in Figure 1A, this mass is sufficient to obtain good glass transition thermal signals for our samples.

Ellipsometry measurements reveal consistent results with the calorimetric experiments. Figure 1B shows the normalized film thickness as a function of temperature for co-deposited m-MTDATA/TPD (50:50) glass mixture at $T_{sub}$=300K. The film thickness for both as-deposited and liquid-cooled glasses increases linearly with temperature due to thermal expansion. A sharp deviation from this linear dependence was observed when the samples started to expand as they transform into a supercooled liquid. In the ellipsometry data, the obtained $T_{onset}$ for co-deposited m-MTDATA/TPD glass is 17.6K higher than the $T_g$ for the liquid-cooled glass, consistent with high kinetic stability. It is expected that the absolute transition temperatures measured by ellipsometry are somewhat lower than those measured by calorimetry due to the lower heating rate employed.

The ratio $T_{onset}/T_g$ is often used to quantify the kinetic stability of vapor-deposited glasses[38]. For the co-deposited m-MTDATA/TPD glasses, calorimetry and ellipsometry measurements show good agreement, with $T_{onset}/T_g$ equal to 1.06 from calorimetric experiments and 1.05 from ellipsometry measurements. Interestingly, $T_{onset}/T_g$ =1.05-1.06 is consistent with the values found in the most stable single-component PVD organic glasses[38]. Furthermore, the $T_{sub}$=300K used to create the ultrastable m-MTDATA/TPD glasses is equal to $0.88T_g$, located in the optimal temperature region (*i.e*, $0.78$-$0.90T_g$) for preparing single-component ultrastable organic glasses[38].

The results above suggest that the surface equilibration mechanism previously advanced for single-component glasses may be applied to understand the kinetic stability of co-deposited m-MTDATA/TPD glasses. To test this hypothesis, we co-deposited m-MTDATA/TPD mixtures with the same mass ratio of 50:50 at five additional substrate temperatures ranging from 210K to 340K. We present the DSC results for these co-deposited glasses in Figure 2A and the results from ellipsometry measurements are shown in Figure S3. As seen from Figure 2A, the $T_{\text{onset}}$ for deposited glasses varies substantially with the substrate temperature used to prepare the glasses. This demonstrates that the kinetic stabilities of co-deposited m-MTDATA/TPD glasses are controlled by the substrate temperatures, consistent with the surface equilibration mechanism.

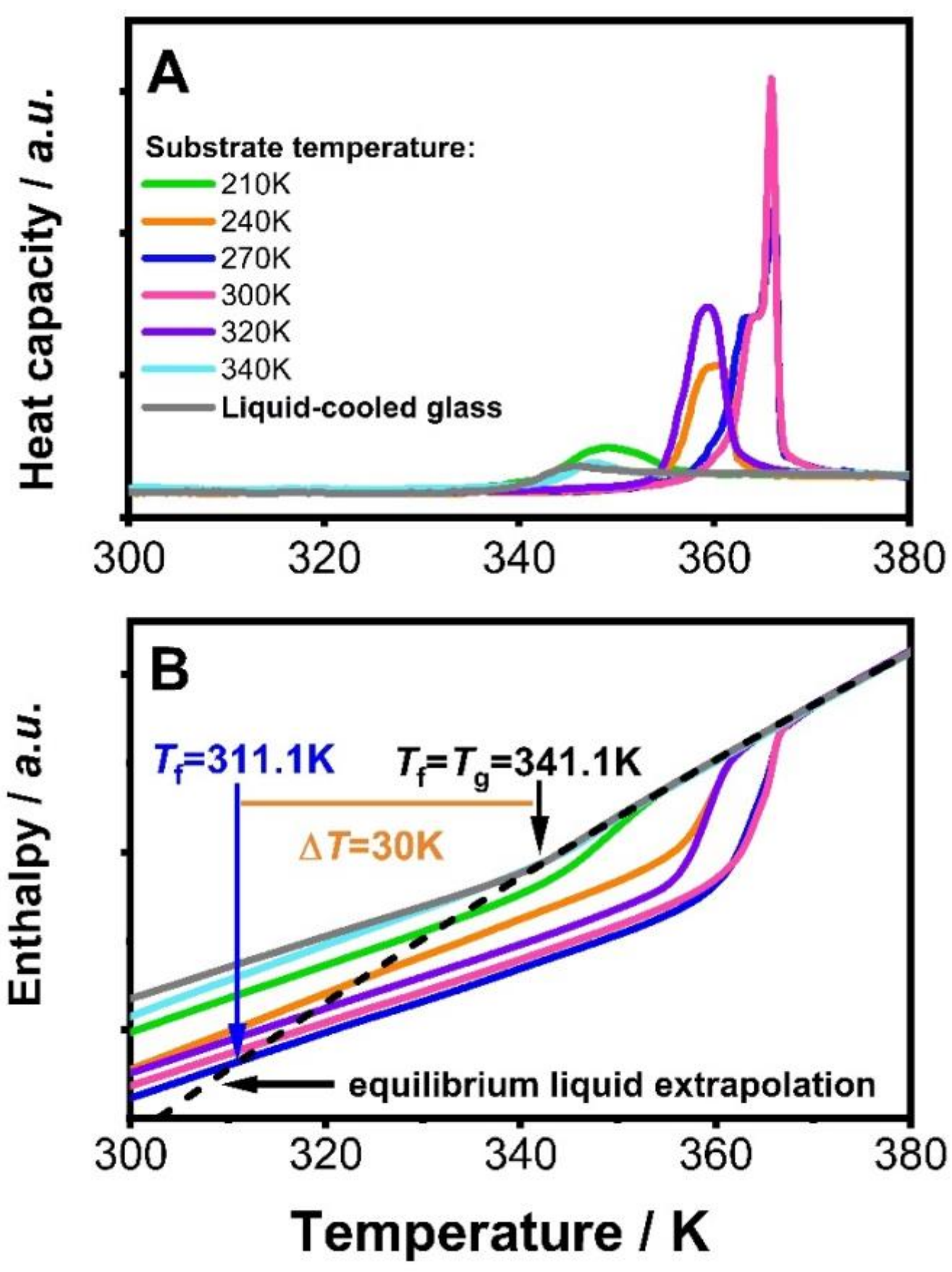

**Figure 2.** A) DSC heating curves for m-MTDATA/TPD mixtures co-deposited at different temperatures. The gray curve denotes the result of the ordinary liquid-cooled glass; B) The enthalpy as a function of temperature for studied m-MTDATA/TPD glasses. The heat capacity of the samples shown in panel A are integrated to obtain the enthalpy data, providing access to the fictive temperature for each sample. The dashed line is the extrapolation of the equilibrium liquid enthalpy to lower temperature by fitting the enthalpy data above $T_g$ (from 345K to 380K) to a quadratic function.

Figure 2A also indicates that the enthalpy for co-deposited m-MTDATA/TPD glasses with the same chemical composition is tunable. One may see that the glass transition endothermic peak area is not constant when the glass mixtures are deposited at different $T_{sub}$. The endothermic peak area quantifies the enthalpy required to transform the glass into the equilibrium liquid state. Figure 2B shows the enthalpy for co-deposited m-MTDATA/TPD glasses as determined by integrating the heat capacity data in Figure 2A. Excluding the glass deposited at $T_{sub}$=340K, the enthalpy of co-deposited glasses is significantly lower than that of the liquid-cooled glass.

The thermodynamic stability of co-deposited m-MTDATA/TPD glasses can be quantitatively compared to single-component PVD glasses using the fictive temperature $T_f$[11]. As shown in Figure 2B, the $T_f$ values for co-deposited m-MTDATA/TPD glasses were determined from the temperature where the as-deposited glass enthalpy matches the (extrapolated) enthalpy data for the equilibrium liquid (the black curve). The m-MTDATA/TPD glass deposited at $T_{sub}$=270K has the lowest $T_f$=311.1K, which is 30K lower than that of the ordinary liquid-cooled glass with $T_f$=$T_g$=341.1K. This result is comparable to the most stable PVD glasses of pure TNB[11] and IMC[39], as well amber glasses aged for tens of millions of years[40,41], whose $T_f$ values are around 30K lower than the glass transition temperature of the liquid-cooled glass.

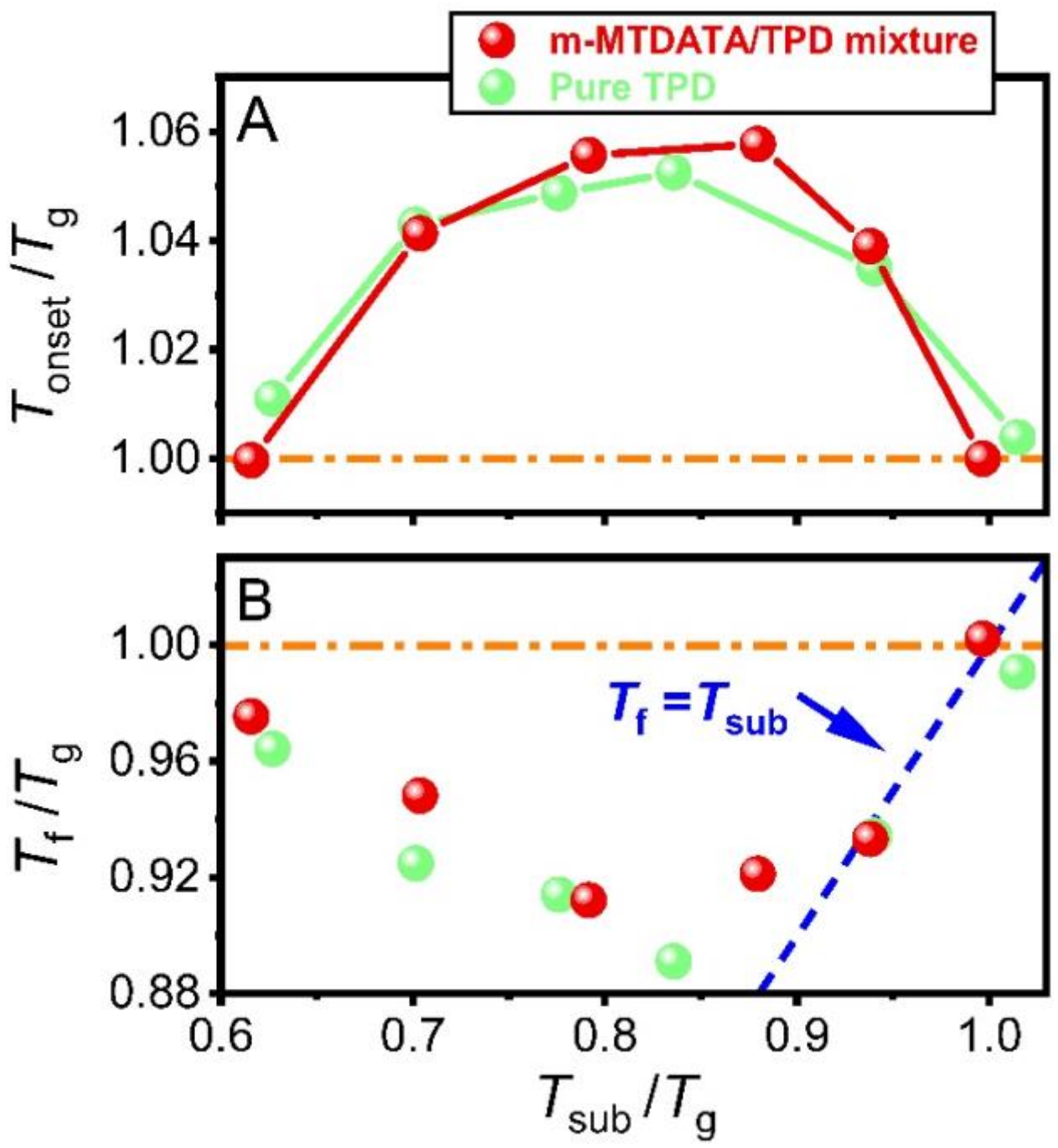

**Figure 3.** A) DSC results of $T_{onset}/T_g$ for deposited m-MTDATA/TPD mixtures (red) and neat TPD (light green) glasses. The solid lines are guides to the eye; B) DSC results of $T_f/T_g$ for deposited m-MTDATA/TPD mixtures (red) and neat TPD (light green) glasses. The blue dashed line represents the $T_{sub} = T_f$ line. The DSC heating curves for as-deposited TPD glasses and the corresponding enthalpy data are displayed in Figure S4.

Figure 3 demonstrates that the kinetic and thermodynamic stability of co-deposited m-MTDATA/TPD glasses are correlated. As shown in this figure, co-deposited m-MTDATA/TPD glasses with higher $T_{onset}/T_g$ (*i.e.*, higher kinetic stability) have lower $T_f/T_g$ (*i.e.*, lower thermodynamic energy state). The highest $T_{onset}/T_g$ and lowest $T_f/T_g$ values are obtained simultaneously when m-MTDATA/TPD glasses are prepared at $T_{sub}$=0.78-0.90$T_g$. These results indicate that the kinetic stability of ultrastable m-MTDATA/TPD glasses are coupled with the occurrence of low energy packing arrangements. Although this behavior has been observed in single-component organic vapor-deposited glasses[9,38,42], it should be noted that this is not a general feature for PVD glasses. For example, the ternary metallic glass of $Zr_{65}Cu_{27.5}Al_{7.5}$ deposited at $0.8T_g$ has an enhanced kinetic stability and elastic modulus, but an enthalpy higher than the corresponding liquid-cooled or annealed glasses[17].

Moreover, the determined trends of $T_{onset}/T_g$ and $T_f/T_g$ as a function of $T_{sub}/T_g$ for co-deposited m-MTDATA/TPD mixtures are in good agreement with those for most organic single-component PVD glasses[9,38,42]. As an example for comparison, we added the $T_{onset}/T_g$ and $T_f/T_g$ values for vapor-deposited neat TPD glasses into Figure 3A and 3B. These values are determined based on the DSC measurements. The corresponding heat capacity and enthalpy data are shown in Figure S4. As seen from Figure 3, the data points of $T_{onset}/T_g$ (or $T_f/T_g$) as a function of $T_{sub}/T_g$ for m-MTDATA/TPD mixtures (red points) exhibit the same pattern as those for pure TPD (light green points). Importantly, similar results are obtained using spectroscopic ellipsometry (see Figure S5).

The non-monotonic dependence of $T_{onset}/T_g$ and $T_f/T_g$ on $T_{sub}/T_g$ revealed in Figure 3 can be understood as a result of the surface equilibration process during deposition. When deposited below $T_g$, there is a thermodynamic driving force to reach the equilibrium liquid state at that temperature. High surface mobility enables molecules to find low energy and high stability configurations before being buried (and immobilized) by further deposition. Direct evidence for this view is presented in Figure 3B; the fictive temperature for m-MTDATA/TPD glasses deposited at $0.94T_g$ is equal to the corresponding substrate temperature (the blue dashed line shows $T_{sub}=T_f$). At lower values of $T_{sub}/T_g$, there will be a larger thermodynamic driving force to form equilibrium state but simultaneously, the surface mobility will decrease. The most stable glasses with the highest $T_{onset}$ (or lowest $T_f$) are formed when high surface mobility is paired with a large thermodynamic driving force. For organic semiconductor compounds deposited at normal rates around 0.1-1 nm/s, this match typically occurs when $T_{sub}$ is around 0.78-0.90$T_g$[36,38]. At lower substrate temperatures than this, the surface mobility is not so high and only moderately stable glasses are formed despite the presence of a larger driving force. Our conclusion that the surface equilibration mechanism explains the stability of co-deposited glasses is consistent with and extends previous work that interpreted molecular orientation in co-deposited glasses using this mechanism[43,44].

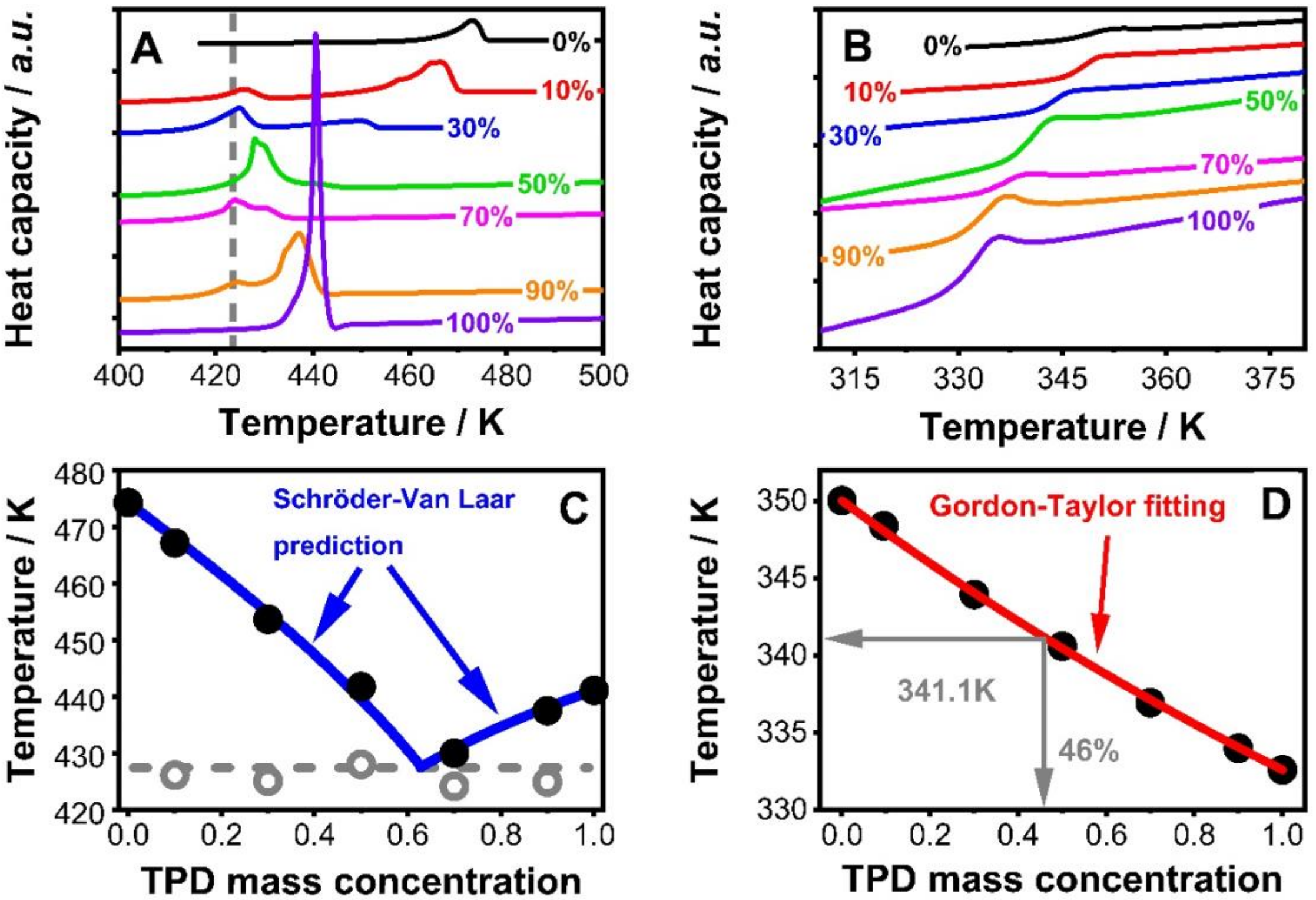

**Figure 4.** A) Differential scanning calorimetry thermograms of the crystalline physical mixtures of m-MTDATA/TPD with different compositions. The percentages denote the corresponding mass concentration of TPD. B) The glass transition region for m-MTDATA/TPD mixtures with different compositions. C) Phase diagram of m-MTDATA/TPD mixtures using experimentally determined data (circles) and a fit to the Schröder-Van Laar equation (blue curves). D) The glass transition temperature of m-MTDATA/TPD mixtures as a function of mass concentration of TPD. The black

circles are experimental data determined from panel B and the red curve is a fit to the Gordon-Taylor equation.

It is expected that the molecular interactions between the two components will have a strong influence on the properties of binary PVD glasses. For example, strong repulsive interaction may lead to component separation during deposition and strong attraction may inhibit surface diffusion. For this reason, we investigated the miscibility of m-MTDATA and TPD in bulk mixtures through calorimetric measurements. Figure 4A illustrates the DSC results of the initial heating process for binary crystalline physical mixtures of m-MTDATA and TPD with different compositions. The percentages give the mass concentration of TPD in each sample. Excluding pure TPD (100% sample) and pure m-MTDATA (0% sample), two melting processes are observed in all mixtures and the lower melting point is independent of the compositions, indicating a eutectic system. Based on these measurements, we constructed the phase diagram of m-MTDATA/TPD mixtures. As seen in Figure 4C, the experimental data is in good agreement with the theoretical prediction from the Schröder-Van Laar equation, indicating that this binary system is quite close to an ideal mixture, which is miscible at any composition.

The glass transitions observed for these physical mixtures provide important checks on miscibility and the composition of the PVD samples. As can be seen in Figure 4B, for each mixture, a single glass transition is observed, demonstrating the formation of a single glassy phase during cooling the molten mixture. According to these DSC data, we determined the glass transition temperature for m-MTDATA/TPD mixtures as a function of TPD concentration. As shown in Figure 4D, the $T_g$ values decrease monotonically with the increase of TPD concentration, and the Gordon-Taylor equation[45] describes the data well. Based on the Gordon-Taylor fitting curve, the co-deposited m-MTDATA/TPD glasses discussed above with $T_g$=341.1K will contain 46% TPD, which is quite close to the TPD concentration of 50% determined based on deposition rate. We infer that the as-prepared PVD glasses are well-mixed, based upon the single heat capacity maximum observed in the DSC experiments (Figure 2A) and the observation that the use of the mixture $T_g$ in Figure 3 produces good correspondence with single component PVD glasses.

Sufficient surface mobility of both TPD and m-MTDATA molecules at temperatures of interest is the key to forming ultrastable co-deposited m-MTDATA/TPD glass mixtures. Figure 5 shows the experimental surface diffusion coefficients ($D_s$) for pure TPD[24] and m-MTDATA[46] plotted as a function of the absolute temperature. The Arrhenius equation was applied to fit the data and extrapolated to lower temperatures. The gray shaded portions of the extrapolated curves are the regions where highly stable TPD or m-MTDATA glasses are formed (using the criterion that $T_{onset}$ $/T_g \geq 1.05$ for TPD and $T_{onset}/T_g \geq 1.04$ for m-MTDATA). Via this procedure, we estimate that the minimum $D_s$ required to form ultrastable neat glasses of TPD and m-MTDATA is around $3\times10^{-24}$ $m^2/s$. Interestingly, the $D_s$ values for both TPD and m-MTDATA are above $3\times10^{-24} m^2/s$ in the temperature region where the most stable co-deposited m-MTDATA/TPD glasses are formed (the pink shaded region). If we assume that $D_s$ for each component in the mixture is not too different from the pure component $D_s$ values, our results can be rationalized by concluding that both components in the co-deposited mixture must have surface mobility above some minimum value if a stable glass is to be formed. When both components have high mobility, and $T_{sub}$ is below the conventional glass transition temperature of the corresponding mixtures, we expect that low energy and high stability packing arrangements can be formed during deposition. From this perspective, a major reason why TPD and m-MTDATA can simultaneously have high mobility at the same temperatures is their comparable glass transition temperatures and ideal mixing. A mixture with strong attractive interactions (non-ideal mixing) might have lower surface mobility that would interfere with stable glass formation, but the nearly ideal nature of the m-MTDATA/TPD mixture rules out this possibility. The arguments in this paragraph are based upon the assumption that surface mobility of one component is not perturbed by the presence of the second component. For stable glass formation, this is the least optimistic scenario. If, for example, both components in a mixture were to have the same (average) surface mobility, then it would not be important that the two components have comparable $T_g$ values. Future work that provides guidance for understanding the surface mobility of multi-component systems would be very useful.

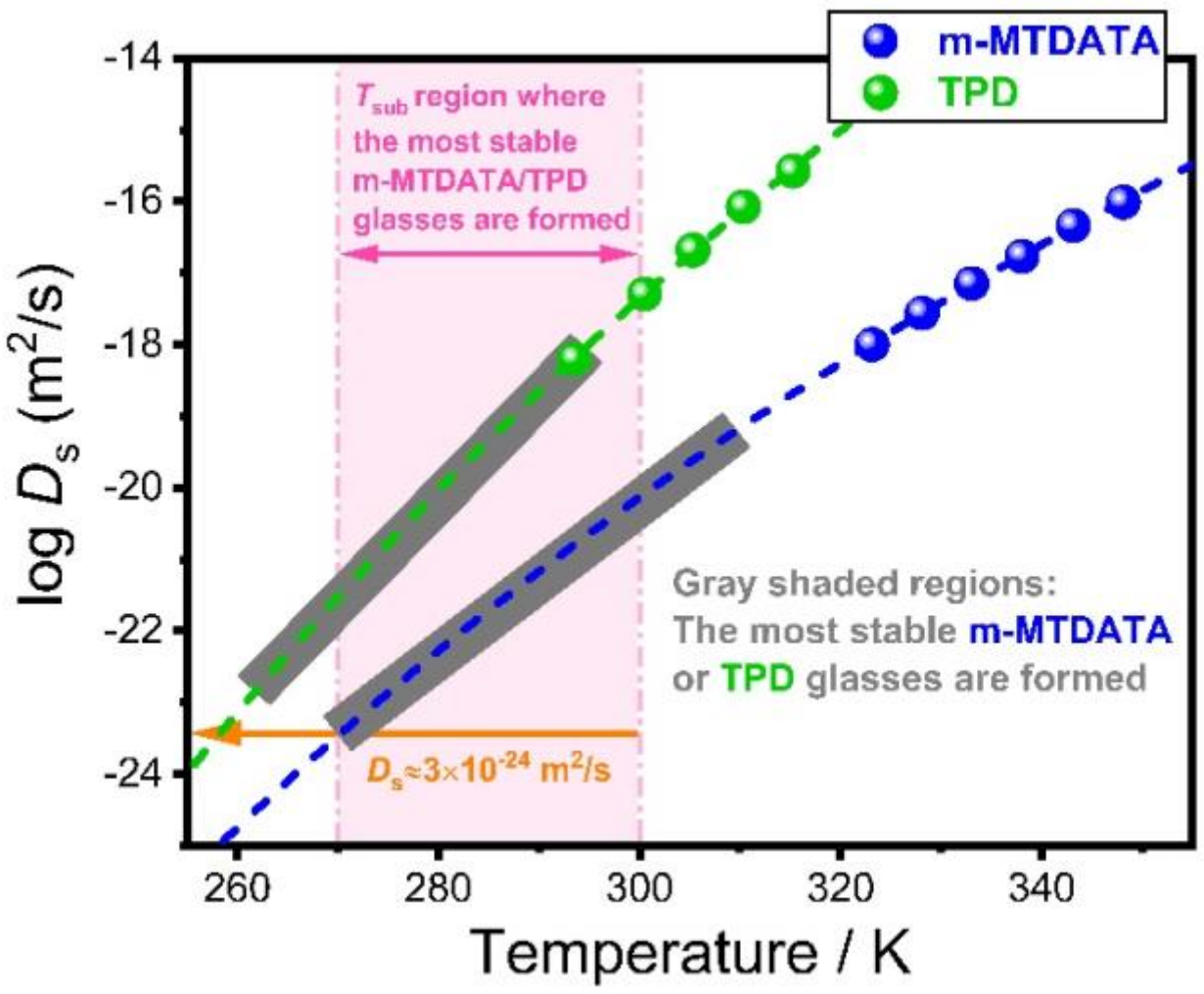

**Figure 5.** Surface diffusion coefficient of pure m-MTDATA (blue) and TPD (green) as a function of absolute temperature. The $D_s$ data for m-MTDATA and TPD were taken from ref.[46] and ref.[24], respectively. The dashed lines denote the Arrhenius extrapolation to predict the $D_s$ values at lower temperatures.

So far, three works have investigated the stability of vapor-deposited organic glass mixtures. In 2013, Whitaker *et al*. reported highly stable glasses of *cis*/*trans*-decalin mixtures across a range of compositions using *in situ* AC nanocalorimetry[28]; the two isomers in this mixture have very similar chemical structures and identical glass transition temperatures. In 2018, Y. Qiu *et al.* showed that PVD produced stable glasses of 5% 4,4'-diphenylazobenzene (DPA) with 95% celecoxib[29]. The current work on m-MTDATA/TPD mixtures considerably expands upon these two papers in using organic semiconductor molecules that have different shapes and $T_g$ values and shows that highly stable glasses are obtained even for 50/50 mixtures. In 2022, M.S. Ki *et al.* reported an ellipsometric study involving the stability of co-deposited organic semiconductors[30]. They reported that co-deposited Liq and BPhen glasses across a wide mixing ratio did not show ultrastable behavior when deposited at $0.80T_g$-$0.89T_g$. We note that M.S. Ki *et al.* also reported that pure Liq failed to form stable glasses via PVD. We see this as a key difference as, for the mixtures studied here, it has been confirmed experimentally that the two pure components (*i.e.*, TPD and m-MTDATA) can form ultrastable glasses individually at proper deposition conditions[36]. Inability of a pure component to form stable glasses via PVD could be interpreted as a lack of surface mobility, which then might explain why mixtures involving that component do not form stable glasses.

Given the large number of organic semiconductor mixtures used as active layers and the importance of their stability in electronic devices, it is useful to consider how general our results may be. We conclude that the surface equilibration mechanism controls the stability of co-deposited m-MTDATA/TPD glasses and suggest that the following three key features are closely related to this conclusion: i) both components can form ultrastable glasses individually via PVD; ii) the components mix well at all compositions without strong association; iii) the glass transition temperature difference between the components is not too large. Since many organic semiconductors can form ultrastable glasses individually when deposited under optimal conditions[36,38,47], based on the knowledge gained here, we expect that other organic semiconductor mixtures can form ultrastable glasses when the chosen compounds have properties similar to m-MTDATA and TPD, including comparable glass transition temperatures, high miscibility, and no strong association. In addition, we anticipate that dilute mixtures of organic semiconductors, which are widely used as light-emitter layers in OLED displays, can form ultrastable glasses when deposited at around $0.85T_g$ as sufficiently dilute solutions are always miscible. This prediction is consistent with the work by J. Ràfols-Ribé *et al.*[34] in which OLEDs containing a co-deposited dilute mixture layer prepared at $0.85T_g$ showed longer device lifetimes, if we assume that these longer lifetimes result from ultrastability.

In summary, our work presents the first case of non-dilute organic semiconductor glass mixtures with exceptional thermodynamic and kinetic stability. We demonstrate that the substrate temperature controls the stability of co-deposited glasses of m-MTDATA and TPD in the same way as it controls the stability of single-component PVD organic glasses. For deposition near $T_g$, the enthalpy equals that expected for the equilibrium liquid. Thus, the surface equilibration mechanism is extended to co-deposited PVD glasses. We suggest that the main reasons why the PVD glasses of this binary system behave like a neat PVD glass are the ideal solution behavior and the comparable surface mobilities over the studied temperature range. In addition, both m-MTDATA and TPD are good glass formers and can individually form ultrastable glasses when deposited under proper conditions. Thus, they can be regarded as a model system to study the properties of binary PVD glasses of organic semiconductors. We expect that other organic semiconductor mixtures can form ultrastable glasses when the chosen compounds have properties similar to m-MTDATA and TPD.

ASSOCIATED CONTENT

**Supporting Information**

The following information is included: Experimental materials and methods; GIWAXS patterns for co-deposited m-MTDATA/TPD film (Figure S1); DSC results for co-deposited m-MTDATA/TPD at $T_{sub}$=270K (Figure S2); Normalized film thickness for co-deposited m-MTDATA/TPD mixtures (Figure S3); DSC results for vapor-deposited TPD (Figure S4); $T_{onset}$ for vapor-deposited m-MTDATA/TPD mixtures and pure TPD (Figure S5).

AUTHOR INFORMATION

**Corresponding author**

*E-mail: ediger@chem.wisc.edu

ORCID:

Shinian Cheng:

Yejung Lee:

Junguang Yu:

Lian Yu:

M. D. Ediger:

**Notes**

The authors declare no competing financial interests.

ACKNOWLEDGMENT

Work by Shinian Cheng, Yejung Lee, and M.D. Ediger was supported by the U.S. Department of Energy, Office of Basic Energy Sciences, Division of Materials Sciences and Engineering, DE-SC0002161. Work by Junguang Yu and Lian Yu was supported by the NSF through the University of Wisconsin Materials Research Science and Engineering Center (Grant No. DMR-1720415).

# Supporting Information for

# Surface Equilibration Mechanism Controls the Stability of a Model Co-deposited Glass Mixture of Organic Semiconductors

*Shinian Cheng*[1], *Yejung Lee*[1], *Junguang Yu*[2], *Lian Yu*[2], *and M. D. Ediger**[,1]

[1]Department of Chemistry, University of Wisconsin-Madison, Madison, Wisconsin, 53706, USA

[2]School of Pharmacy, University of Wisconsin-Madison, Madison, Wisconsin, 53705, USA

**Corresponding Author**

*Email: ediger@chem.wisc.edu

This supporting information file includes:

Description of Experimental materials and methods

Figure S1: GIWAXS patterns for co-deposited films.

Figure S2: DSC results for co-deposited m-MTDATA/TPD at $T_{sub}$=270K.

Figure S3: Normalized film thickness for co-deposited m-MTDATA/TPD mixtures.

Figure S4: DSC results for vapor-deposited TPD glasses.

Figure S5: $T_{onset}$ for vapor-deposited m-MTDATA/TPD mixtures and pure TPD.

References

## Materials and Methods

**Vapor-deposited glassy film preparation.** TPD (99% purity) and m-MTDATA (98.7% purity) were purchased from Sigma-Aldrich and used without further purification. Physical vapor-deposited glasses were prepared in a high-vacuum chamber with a base pressure ~$10^{-6}$ Torr. For co-deposition, the deposition rate for each component was controlled by individually heating two independent crucibles using resistive wire heaters to achieve the desired 50:50 glassy mixtures. For all depositions, the rate was 0.42±0.03nm/s and monitored using a quartz crystal microbalance (QCM). Vapor-deposited glassy films with a thickness of 1200-1400nm for differential scanning calorimetry measurements were deposited onto 120nm thick gold foil (purchased from Barnabas Gold). Films with a thickness of 380-400nm for spectroscopic ellipsometry and GIWAXS measurements were deposited onto one-side polished silicon wafers (purchased from Virginia Semiconductor). The film thickness is measured by QCM and corrected using spectroscopic ellipsometry. The substrate temperature was held constant during deposition using a Lakeshore controller with platinum RTD sensors.

**Bulk mixture preparation.** Crystalline physical mixtures of TPD and m-MTDATA binary systems were prepared by mixing and grinding using a mortar and pestle. Seven samples with mass concentrations of TPD were prepared: 0%, 10%, 30%, 50%, 70%, 90%, and 100%.

**Differential scanning calorimetry (DSC) measurements.** Thermal analysis of bulk and vapor-deposited samples were performed using a TA Q2000 differential scanning calorimeter (New Castle, DE). To determine the miscibility of TPD and m-MTDATA and the glass transition temperatures of m-MTDATA/TPD mixtures, a sealed aluminum DSC pan with around 5mg crystalline physical mixtures was loaded into the instrument. To determine the thermal stability of PVD glasses, the vapor-deposited glassy films with the attached gold foil (120nm thickness) were folded and loaded into a Tzero pan; the pan was sealed by a Tzero lid using a crimper press to achieve good contact between the tested sample and the pan. A similar measurement protocol was used previously by Rodríguez-Viejo *et al.*[1]. For all DSC measurements, the rate is 10K/min for both heating and cooling processes under 50mL/min $N_2$ purge. In this paper, $T_g$ denotes the glass transition temperature of liquid-cooled glasses for either mixtures or pure samples.

**Spectroscopic ellipsometry measurements.** The thickness and optical properties of glassy films were measured using a spectroscopic ellipsometer (J.A. Woollam M-2000U) with a custom-built temperature control and translation stage. We first mapped each sample at room temperature with variable angle measurements and then performed temperature scanning measurements with 1K/min at a fixed incidence angle of 70°. Three temperature cycles were performed for each sample. The collected optical parameters in the wavelength range of 600-1000nm were fitted to a three-layer model consisting of the silicon substrate layer, 2nm thick native oxide layer, and the transparent sample layer. The anisotropic Cauchy model previously discussed[2] was applied to determine the thickness evolution of the sample layer with temperature and thereby the $T_{onset}$ for as-deposited glasses and $T_g$ for liquid-cooled glasses.

**Grazing-Incidence Wide-Angle X-ray Scattering (GIWAXS).** GIWAXS for co-deposited films was performed at beamline 11-3 with x-ray energy 12.7 keV at the Standford Synchrotron Radiation Lightsource (SSRL). The images were acquired at an incidence angle 0.14° (above the critical angle for studied materials) with the exposure time 90s. The sample to detector distance is 301.2 mm and calibrated with $LaB_6$. Scans were performed in a helium atmosphere.

## GIWAXS patterns for co-deposited m-MTDATA/TPD films

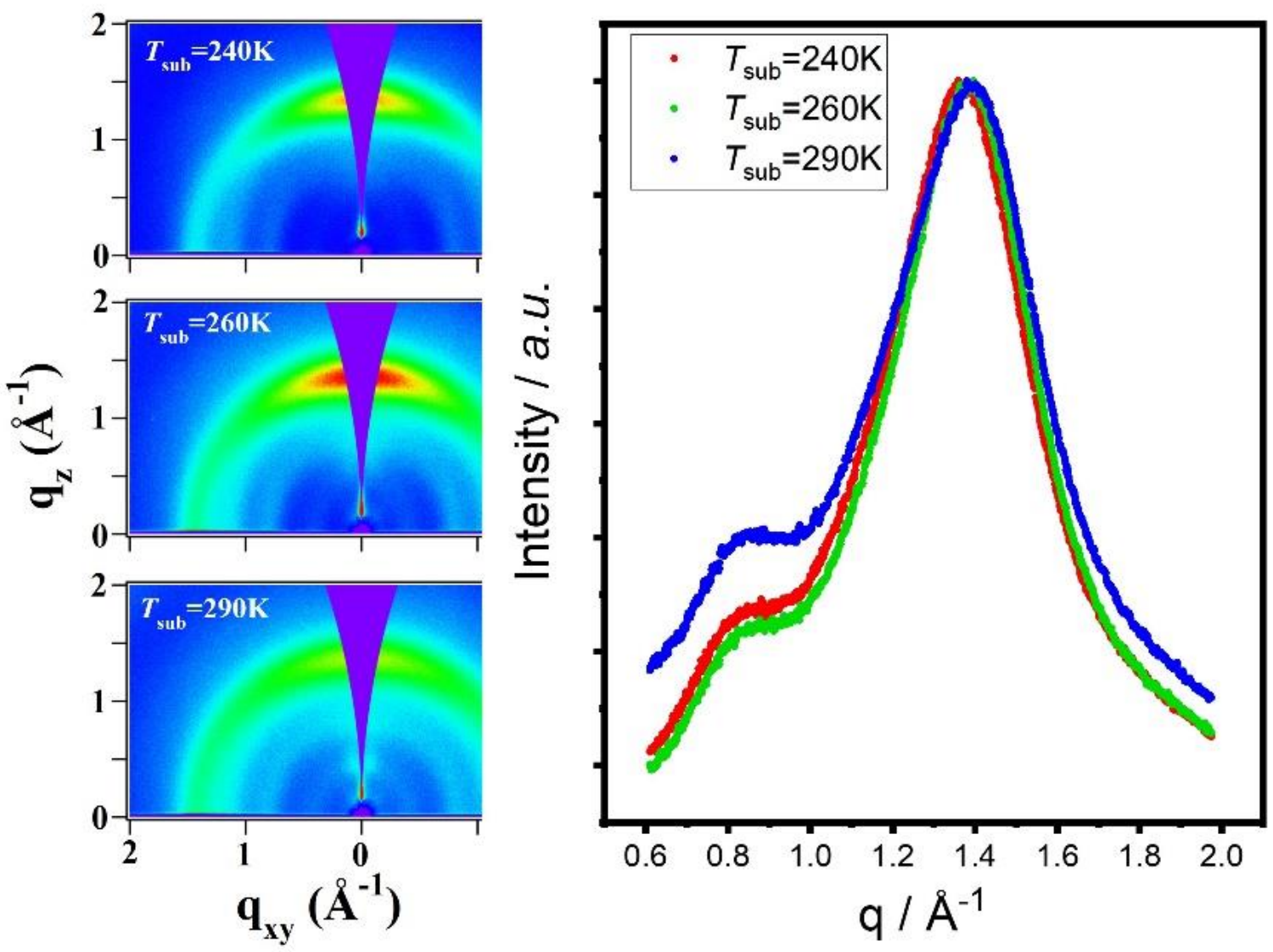

**Figure S1.** GIWAXS patterns for co-deposited m-MTDATA/TPD films at $T_{sub}$=240 K, 260 K, and 290 K. The scattering patterns were corrected for the scattering geometry, which results in the missing wedge along $q_z$. The broad GIWAXS scattering features indicate that the as-deposited films studied here are not crystalline.

## DSC results for co-deposited m-MTDATA/TPD at Tsub=270K

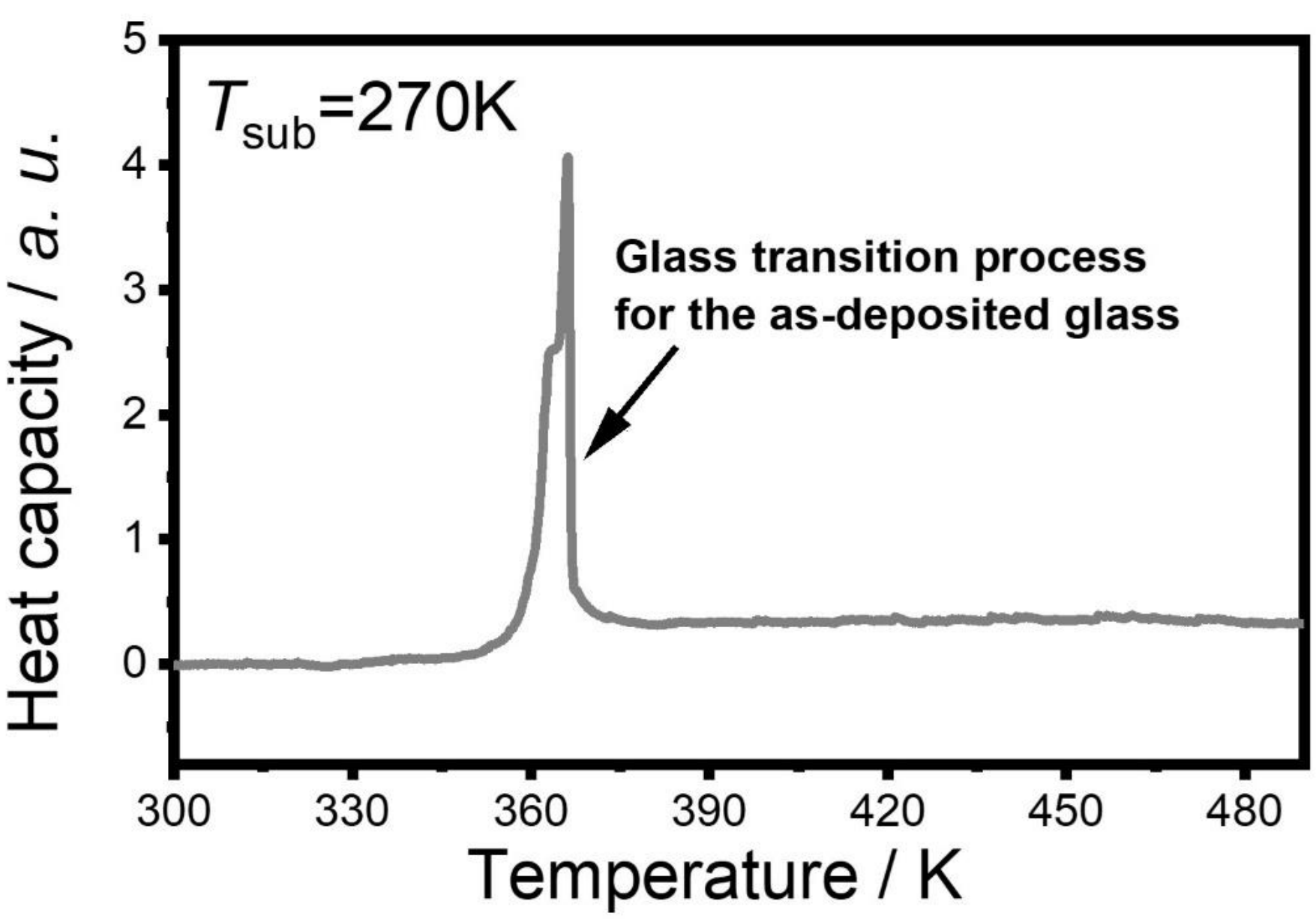

**Figure S2**. DSC results for as-deposited m-MTDATA/TPD film at $T_{sub}$=270K in temperature range from 300K to 490K. Only a glass transition process for the as-deposited sample is observed.

## Normalized film thickness for co-deposited m-MTDATA/TPD mixtures

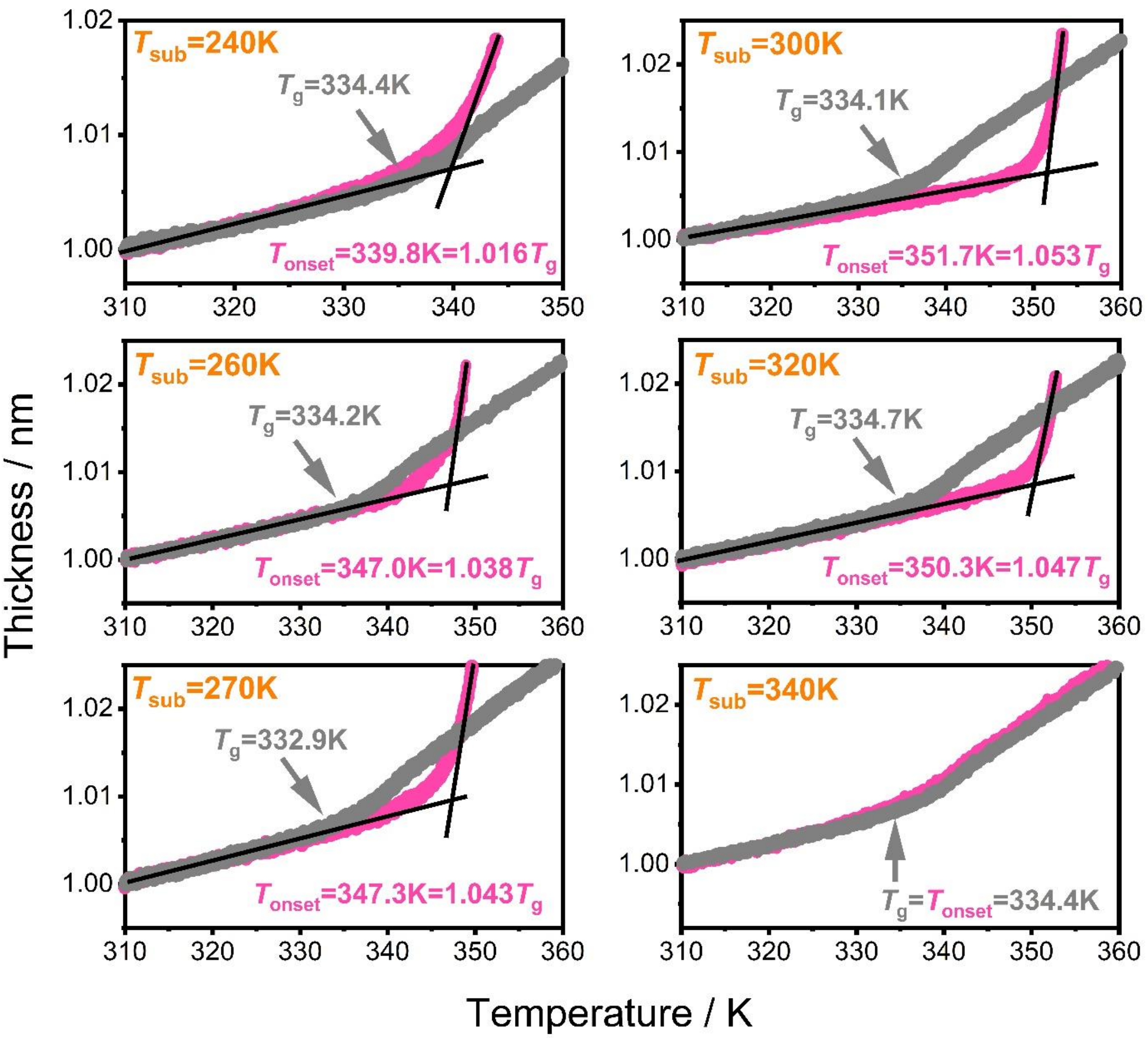

**Figure S3.** Normalized film thickness at 310K as a function of temperature for co-deposited m-MTDATA/TPD mixtures at different substrate temperatures presented in each graph. The data are determined through ellipsometry ramping experiments. Pink data is collected in the first heating process for the as-deposited glasses, while gray data is obtained in the second heating run for the corresponding liquid-cooled glasses. The heating and cooling rate was 1K/min.

## DSC results for vapor-deposited TPD glasses

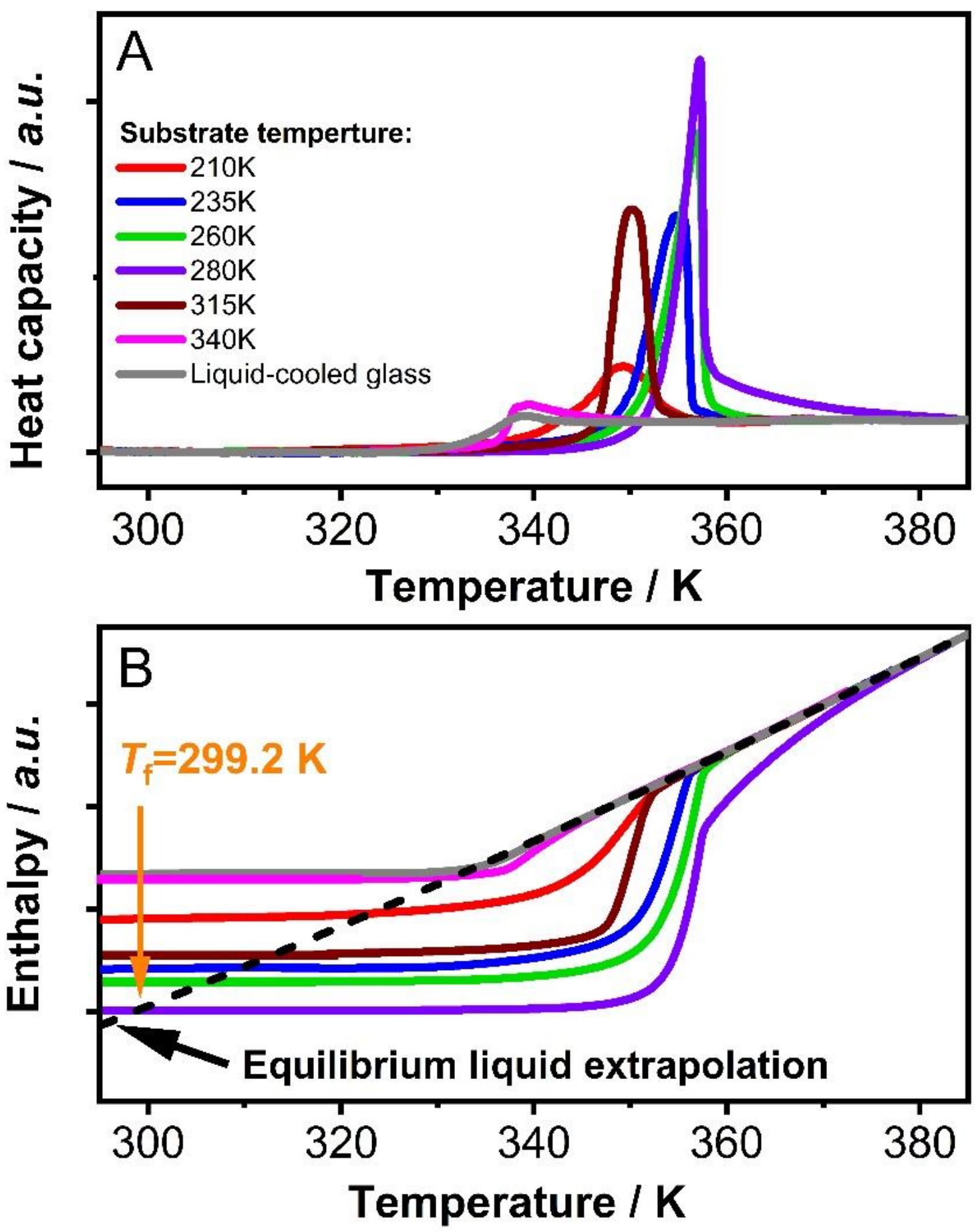

**Figure S4.** A) DSC heating curves for pure TPD glasses deposited at different substrate temperatures. The gray curve denotes the DSC result of the ordinary liquid-cooled TPD glass; B) The enthalpy as a function of temperature for deposited TPD glasses. The heat capacity of the samples shown in panel A are integrated to obtain the enthalpy data.

## $T_{onset}$ for vapor-deposited m-MTDATA/TPD mixtures and pure TPD

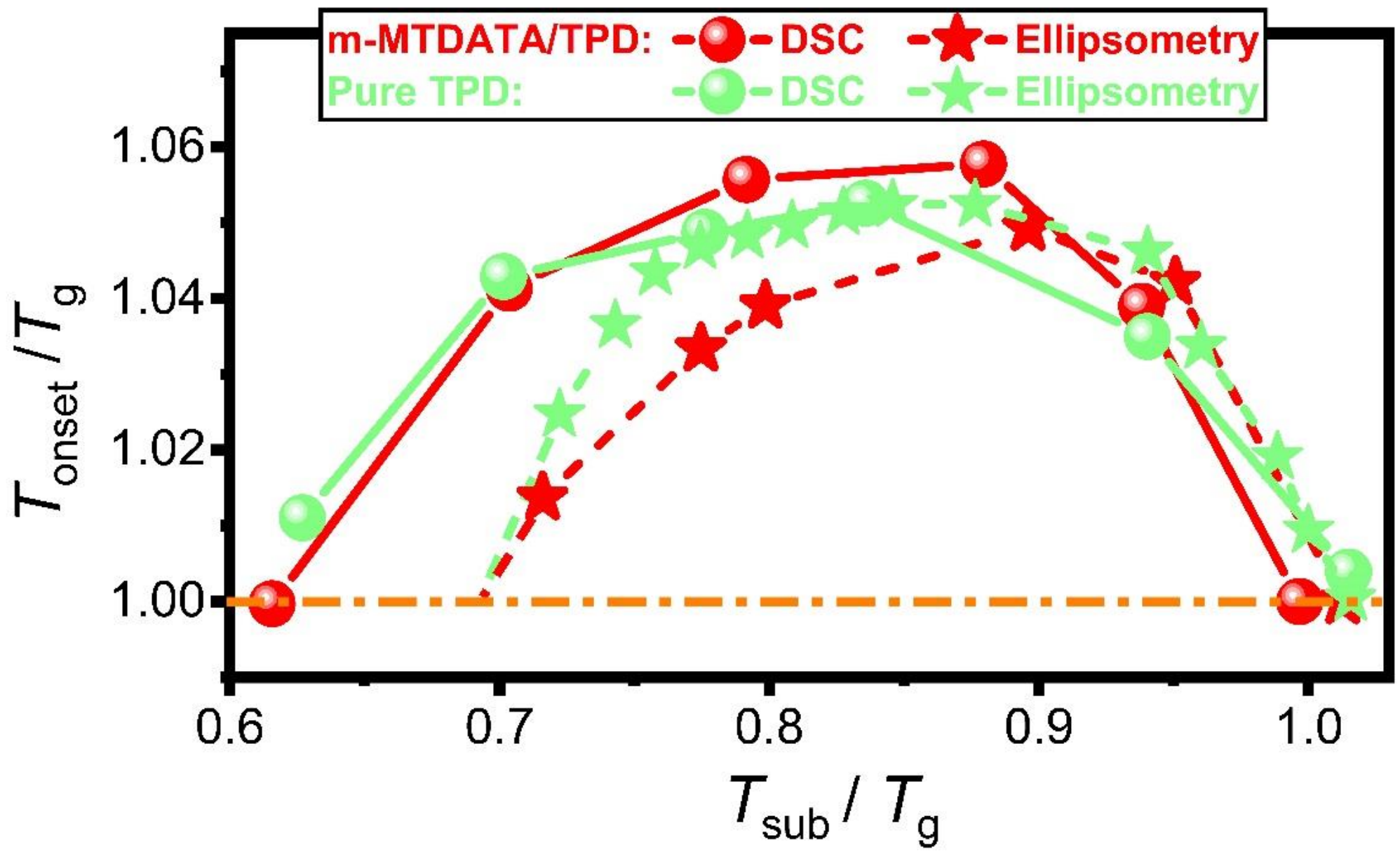

**Figure S5.** The $T_{onset}/T_g$ values for m-MTDATA/TPD mixtures (red) and pure TPD (light green) vapor-deposited at different $T_{sub}/T_g$. The solid and dashed lines are a guide to the eye. For pure TPD, the ellipsometry $T_{onset}/T_g$ data is taken from ref.[3], and the DSC dataset is determined from Figure S4.